\documentclass[manuscript]{aastex}
\usepackage{epsfig, graphicx}

\slugcomment{Submitted to Ap. J. Letters}

\shorttitle{The progenitor and evolution of SN 2009kr}
\shortauthors{Fraser et al.}

\begin{document}


\title{On the progenitor and early evolution of the type II supernova 2009kr}


\author{M. Fraser}
\affil{Astrophysics Research Center, School of Mathematics and Physics, Queens University Belfast, BT7 1NN, UK}

\and

\author{K. Tak\'ats\altaffilmark{1}}
\affil{Department of Optics and Quantum Electronics, University of Szeged, D\'om t\'er 9., H-6720, Szeged, Hungary}

\and

\author{A. Pastorello, S.J. Smartt}
\affil{Astrophysics Research Center, School of Mathematics and Physics, Queens University Belfast, BT7 1NN, UK}

\and

\author{S. Mattila}
\affil{Tuorla Observatory, Department of Physics and Astronomy, University of Turku, V\"{a}is\"{a}l\"{a}ntie 20, FI-21500, Finland}

\and

\author{M-T. Botticella, S. Valenti}
\affil{Astrophysics Research Center, School of Mathematics and Physics, Queens University Belfast, BT7 1NN, UK}

\and

\author{ M. Ergon, J. Sollerman}
\affil{Oskar Klein Centre, Department of Astronomy, AlbaNova, Stockholm University, 106 91 Stockholm, Sweden}

\and

\author{S. Benetti, F. Bufano}
\affil{INAF, Osservatorio Astronomico di Padova, Vicolo dell'Osservatorio 5, IT 35122 Padova, Italy}

\and

\author{R.M. Crockett}
\affil{Oxford Astrophysics, Department of Physics, Denys Wilkinson Building, Keble Road, Oxford, OX1 3RH, UK}

\and

\author{I.J. Danziger}
\affil{INAF Osservatorio Astronomico di Trieste, Via G. B. Tiepolo 11, IT 34131 Trieste, Italy}

\and

\author{J.R. Maund}
\affil{Dark Cosmology Centre, Niels Bohr Institute, University of Copenhagen, 2100 Copenhagen, Denmark}

\and

\author{S. Taubenberger}
\affil{ Max-Planck-Institut f\"ur Astrophysik, Karl-Schwarzschild-Str. 1, 85741 Garching bei M\"unchen, Germany}

\and

\author{M. Turatto}
\affil{INAF, Osservatorio Astronomico di Catania, Via S.Sofia 78, 95123 Catania, Italy}

\altaffiltext{1}{Astrophysics Research Center, School of Mathematics and Physics, Queens University Belfast, BT7 1NN, UK}

\email{mfraser02@qub.ac.uk}

\begin{abstract}
We report the identification of a source coincident with SN 2009kr in HST pre-explosion images. The object appears to be a single point source with an intrinsic colour $V-I$=1.1 and $M_V = -7.6$. If this is a single star it would be a yellow supergiant of $\log L/L_{\odot} \sim 5.1$ and a mass of $15^{+5}_{-4}$ M$_{\odot}$. The spatial resolution does not allow us yet to definitively determine if the progenitor object is a single star, a binary system, or a compact cluster. We show that the early lightcurve is flat, similar to IIP SNe, but that the the spectra are somewhat peculiar, displaying unusual P-Cygni profiles. The evolution of the expanding ejecta will play an important role in understanding the progenitor object. 
\end{abstract}


\section{Introduction}
\label{s1}

In recent years the hypothesis that red supergiants between 8 and 17 solar masses explode as Type II Plateau (IIP) SNe has been confirmed by several direct detections of progenitors in pre-explosion images, and by setting upper mass limits from non-detections (see, for example, Smartt \citealp{Sma09a}). Two progenitor candidates which show Luminous Blue Variable characterics, either in quiescence or outburst, have been found for SNe with clear signatures of circumstellar interaction, the Ibn 2006jc and IIn 2005gl (Pastorello et al. \citealp{Pas07}; Gal-Yam \& Leonard \citealp{Gal09}). Despite these successes, the total number of supernovae for which we have definite progenitor identifications is still low. \object{SN 2009kr} was found in the spiral galaxy NGC 1832 on Nov. 6 2009 by Itagaki (Nakano \citealp{Nak09}), at an unfiltered magnitude of 16.0. Tendulkar et al. \cite{Ten09} obtained a spectrum on Nov. 8, and suggested that SN 2009kr showed the features of a IIn supernova. The authors reported a blue, featureless continuum, with narrow hydrogen Balmer emission lines and a H$\alpha$ line-width of 880 km s$^{-1}$. A second spectrum was obtained by Steele et al. \cite{Ste09} on Nov. 9, which showed weak P-Cygni absorption in the hydrogen lines. Based on this, and indications that the narrow Balmer lines seen by Tendulkar et al. were in fact produced by a nearby HII region, Steele et al. claimed SN 2009kr to be a young Type IIP.

In this paper, we have taken the host galaxy of SN 2009kr to be at a distance of $26.2\pm1.8$ Mpc (from the recessional velocity, correcting for Virgo-centric infall; with values taken from NED), which corresponds to a distance modulus $(m-M)$ of $32.09\pm 0.15$ mag. Estimates from the Tully-Fisher relation however, give a significantly higher value of $(m-M)=32.61\pm0.43$ mag (Willick et al. \citealp{Wil97}). This value may be unreliable as NGC 1832 has a faint absolute magnitude, where the T-F relation has a larger scatter. Terry et al. \cite{Ter02} give a low value of $(m-M) = 30.76 \pm 0.41$ mag from the method of ``sosie'' galaxies. These two methods bracket our recessional velocity distance; in the absence of any indication of which method is the most reliable in this case, we used the recessional velocity based distance as a compromise, with a conservative error of $\pm 0.5$ mag to reflect the uncertainty. We have used the standard Schlegel et al. \cite{Sch98} relation for Galactic extinction, giving $A_V = 0.242$ and $A_I = 0.142$ mag (values taken from NED). Host extinction is assumed to be negligible, as our spectra of the supernova appear blue and lack prominent NaI D lines. We have used the de-projected galactocentric distance (5.6 kpc) of the supernova to estimate the metallicity of the region in which SN 2009kr exploded. We first determined a characteristic oxygen abundance for the host galaxy of 12 + log(O/H) = $8.33 \pm 0.24$ dex from its absolute B magnitude (Pilyugin et al. \citealp{Pil04}). Using the radial oxygen abundance gradient taken from Boissier \& Prantzos \cite{Boi09}, $-0.50$ dex / R$_{25}$, we find a metallicity of 12 + log(O/H) = $8.06 \pm 0.24$ dex. While this is low compared to the location of other Type IIP SNe (Smartt et al. \citealp{Sma09b}), we note that empirical metallicity calibrations suffer considerable uncertainties (see for example, Bresolin \citealp{Bre08}).

\section{Observations and data analysis}

\begin{figure}
\plotone{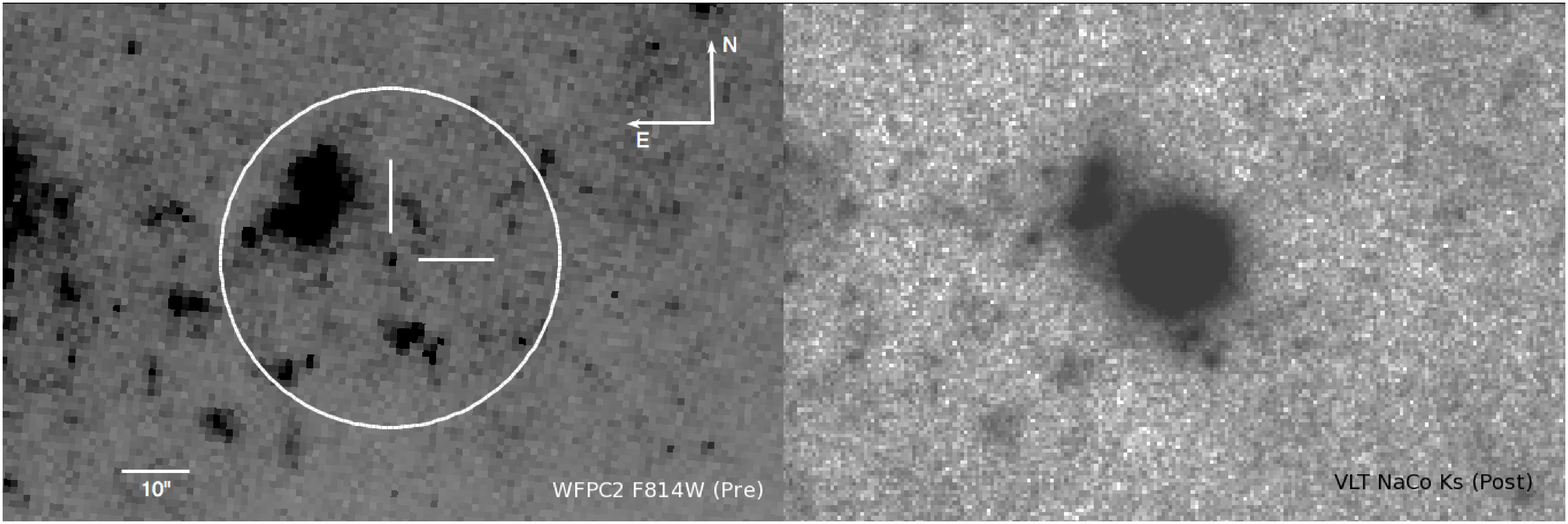}
\caption{Pre- and post- explosion images of the site of SN 2009kr. Each image is centered on the supernova location, with the WFPC2 image being transformed to the same scale and orientation as the NIRI post-explosion image, as per the text. Cross marks have been added to indicate the progenitor location, circle shows location of sources in Fig. \ref{fig2}. Progenitor candidate is located at WFPC2 pixel coordinates (432.8, 149.6).
\label{fig1}}
\end{figure}

The host galaxy of SN 2009kr was observed on 2008 Jan 11 ($\sim 660$ days before explosion) with the Wide Field and Planetary Camera 2 (WFPC2) onboard the Hubble Space Telescope as part of HST Program 10877. Data were reduced and calibrated by the on-the-fly calibration pipeline, and downloaded from the MAST archive at STScI\footnote{http://archive.stsci.edu/}. Pre-explosion images consisted of two 230s exposures in the F555W filter, and two 350s exposures in the F814W filter. Each pair of images was combined with the {\sc crrej} routine within {\sc iraf}\footnote{http://iraf.noao.edu} to remove cosmic rays. The site of SN 2009kr fell on the WF3 chip, which has a pixel scale of 0.1 $\arcsec$/pixel. A 2340s on-source integration in the Ks filter of SN 2009kr was obtained with Naos-Conica (NaCo) on the VLT on 2009 Nov. 21 as part of our supernova progenitor program\footnote{Based on observations collected at the European Organisation for Astronomical Research in the Southern Hemisphere, Chile, Programs 083.D-0131 \& 184.D-1140}. The S54 camera was used (pixel scale of 0.054$\arcsec$) across a  56$\arcsec$ x 56$\arcsec$ field of view and the supernova itself ($m_V \sim 15$ mag) was used as a natural guide star to provide adaptive optics correction for the image. Data were reduced using standard {\sc iraf} routines. The pre-explosion WFPC2 image and post-explosion NaCo images are shown in Fig \ref{fig1}.

To determine the position of SN 2009kr in the pre-explosion images, 18 sources common to both frames were identified. Their centroids were measured with aperture photometry and the resulting list of matched coordinates was used to determine a geometrical transformation between the pre- and post-explosion images with {\sc iraf geomap}. Translations and independent rotation and scaling in x and y were allowed for. Six outliers lying more than one pixel outside the fit were rejected and the fit re-calculated. The final rms error in the transformation (as taken from the output of {\sc geomap}) was 34 mas. The position of the supernova was measured using the three different centering algorithms within {\sc phot} (gaussian, centroid and ofilter) with the mean of the three results being taken as the position and the standard deviation of the results as the uncertainty (39 mas). The same procedure was used to determine the position and uncertainty (3 mas) in the progenitor candidate. The position of the supernova as measured in the post-explosion NaCo image was then transformed into WFPC2 coordinate system using the {\sc geoxytran} task and the transformation determined previously. The uncertainties in the progenitor candidate and supernova positions were added in quadrature together with the uncertainty in the transformation to give the total uncertainty in the procedure. The separation between the progenitor candidate and the supernova was found to be 6 mas, which is well within the total uncertainty of 52 mas. We thus conclude that the source indicated in Fig. \ref{fig1} is coincident with SN 2009kr. Li et al. \cite{Li09} identified the same progenitor in the archival WFPC2 images, using an alignment to a post-explosion image obtained with the CFHT and Mega-Cam.

To characterise the progenitor, we carried out PSF-fitting photometry with the {\sc hstphot} package (Dolphin \citealp{Dol00}). The coincident source was clearly detected in both the F814W and F555W images by {\sc hstphot} at the 10 $\sigma$ level in F555W and 14 $\sigma$ in F814W. We ran {\sc hstphot} twice, first measuring the sky at each pixel from the mean of its neighboring pixels, and secondly recalculating the local sky at the location of the progenitor using the pixels immediately outside the photometric aperture. Measured magnitudes were $m_V = 24.53 \pm 0.11$ and $m_I = 23.47 \pm 0.08$ for the first run, and $m_V = 24.71 \pm 0.14$ and $m_I = 23.48 \pm 0.10$ for the second. $m_I$ is unchanged within the uncertainties, however there is a 0.2 magnitude difference in $m_V$. We have taken the magnitudes from the second run as more reliable, as the recalculated sky value is likely more appropriate for backgrounds that vary rapidly over short distances, such as in this case. To give a more reliable uncertainty estimate for our photometry, we have added the error in the output of {\sc hstphot} in quadrature with the difference between the magnitudes given by the first and second runs. This gives a final progenitor magnitude of $m_V = 24.71 \pm 0.23$ and $m_I = 23.48 \pm 0.10$ from {\sc hstphot}. With the distance modulus (and its uncertainty) and extinction from Sect. \ref{s1}, this corresponds to an absolute magnitude of $M_V = -7.62 \pm 0.55$ and $M_I = -8.75 \pm 0.51$, and $V-I  = 1.13\pm0.25$.

As an independent check of the output of {\sc hstphot}, aperture photometry was performed. We applied the same CTE corrections as {\sc hstphot}; aperture corrections were determined from bright isolated sources. The magnitudes obtained from aperture photometry are 0.2 magnitudes brighter than those given by {\sc hstphot}; we attribute this as likely due to extra flux from nearby sources which the PSF-fitting of {\sc hstphot} can better remove. We obtain $V-I = 1.03$, which agrees with {\sc hstphot} within the uncertainties.

The $\chi^2$ and sharpness statistics in the output of {\sc hstphot} suggest the source is a single star-like PSF; we note however that at the distance of NGC 1832 a single WF chip pixel corresponds to $\sim$ 13 pc. To attempt to constrain the nature of the progenitor further, we used {\sc ishape} (Larsen \citealp{Lar99}) to characterize the best PSF fit to the progenitor source. {\sc ishape} is intended to determine the true shape and size of sources down to 10\% of the PSF size, for sources with a S/N of $\sim30$ or greater. As our progenitor is detected at a much lower significance level than this, we are wary of relying on the results of {\sc ishape} in this case. This caveat notwithstanding, we used the {\sc psf} task within the {\sc iraf daophot} package to create a PSF based on eight bright, isolated sources on the WF3 chip in the F814W filter. This PSF was then used within {\sc ishape} to fit the progenitor. The low $\chi^2$ of the fit indicates that the source is not clearly extended.

Bastian et al. \cite{Bas05} suggest that sources with $M_V < -8.6$ are likely to be clusters. While the source detected is $\sim$ 1 magnitude fainter than this, further extinction in the host galaxy could increase the true absolute magnitude of the progenitor candidate. In Fig. \ref{fig2} we plot a colour-magnitude diagram (CMD) of 21 sources detected with a S/N of 5 $\sigma$ or above in both filter pre-explosion images. We used the fitting statistics in the output of {\sc hstphot} to try to identify extended sources. While {\sc hstphot} should return $\chi^2 < 1.5$ and $-0.3 <$ sharpness $< 0.3$ for single, uncrowded sources, values of $\chi^2 < 2.5$ and $-0.5 <$ sharpness $< 0.5$ are reasonable. While all sources within 2.5$\arcsec$ were within the sharpness limits, four sources were found to have  $1.5 < \chi^2 < 2.5$, and have been classed as possibly extended on this basis, this is also supported by visual inspection of the images. We also inspected pre-explosion R-band and $H\alpha$ images taken with the Wide Field Camera (WFC) on the 2.5m INT on 20 Mar 2005. The continuum subtracted $H\alpha$ image was aligned to the pre-explosion WFPC2 image, and although the pixel scale of WFC (0.33$\arcsec$/pix) is poor compared to WFPC2, identifiable sources of $H\alpha$ emission are visible. Four of the five brightest $V$-band sources are associated with $H\alpha$ emission, indicating they are young stellar clusters. There is no obviously strong $H\alpha$ emission at the position of SN 2009kr in these images.

\section{Discussion}

The $V-I$ colour of our progenitor after correcting for foreground extinction is 1.13 $\pm 0.25$ mag. Taking the intrinsic colours of supergiants from Drilling \& Landolt \cite{Dri00}, this corresponds to a spectral type of G6, with an effective temperature of 4850K, although the uncertainty in colour means it could be as early as G1 or as late as K2. For a G6 star, we take a V-band bolometric correction of  $-0.36 \pm 0.2$ mag from Drilling \& Landolt, with the uncertainty corresponding to the variation in bolometric corrections across the range of possible spectral type. Applying the bolometric correction to the absolute magnitude of the progenitor we find the bolometric magnitude to be $-8.0 \pm 0.6$. Using the standard relation between bolometric magnitude and luminosity
\begin{equation}
log \frac{L(T_{eff})}{L_{\odot}} = \frac{M_{bol}-4.74}{-2.5}
\end{equation}
we find a progenitor luminosity of log $L / L_{\odot} = 5.10 \pm 0.24$. We show this luminosity and temperature on a H-R diagram in Fig. \ref{fig3}, together with the {\sc stars} evolutionary tracks of Eldridge \& Tout \cite{Eld04a}. We have plotted tracks for two different metallicities, Z = 0.004 and Z = 0.008 (comparable to the SMC and LMC respectively), we note however that the final luminosity of each pair of metallicities is extremely close, indicating that the precise progenitor metallicity is not a significant source of error in determining the progenitor mass. While the progenitor lies closest to the $20 M_{\odot}$ track, it is important to remember that a $20 M_{\odot}$ model will explode when it has an Fe core, and hence its luminosity will be higher by about 0.25 dex. It is more appropriate to compare the progenitor luminosity to the luminosity reached by the models at their end point, which corresponds to the helium core luminosity at the end of core carbon burning, i.e. the initial mass - final luminosity diagram as discussed in Smartt et al. \cite{Sma09b}. We hence find a progenitor mass of $15^{+5}_{-4}$ M$_{\odot}$.

\begin{figure}
\includegraphics[angle=270,scale=0.60]{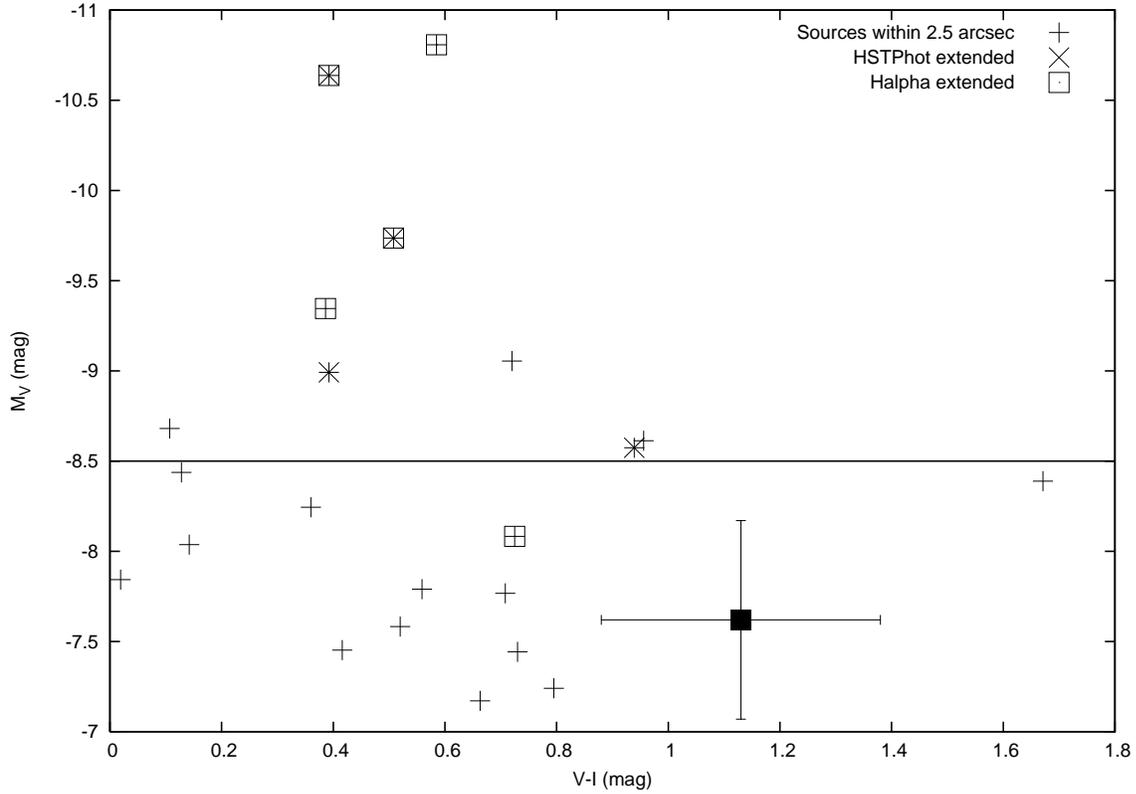}
\caption{CMD of all WFPC2 sources detected at the 5 $\sigma$ level or higher in both filters, within an arbitrary radius of 2.5'' centered on the progenitor location. Sources which have indications that they may be extended from either $H_{\alpha}$ or the output of {\sc hstphot} have been marked with a square or cross respectively. The dashed line at $M_V = -8.5$ is an indication of the threshold beyond which all sources are likely clusters. Progenitor is marked with error bars.
\label{fig2}}
\end{figure}

\begin{figure}
\plotone{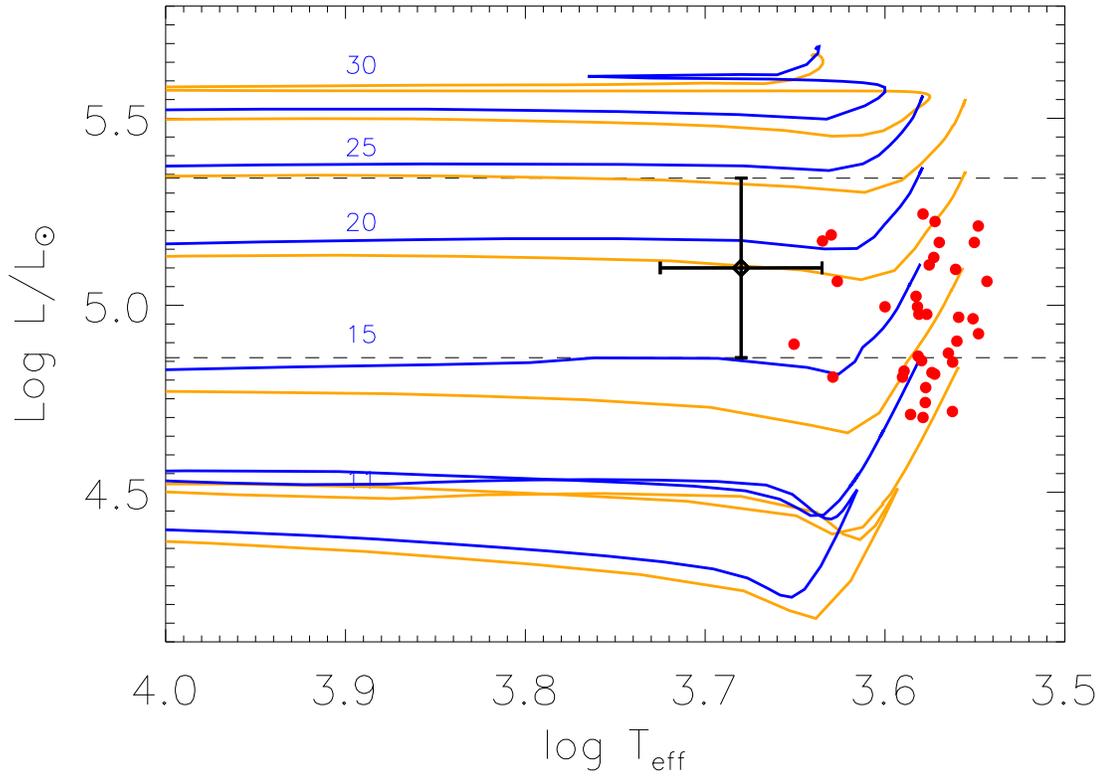}
\caption{HR diagram showing evolutionary tracks and the progenitor of SN 2009kr. The orange tracks correspond to Z=0.008, while the blue tracks have a metallicity of Z=0.004. Red dots correspond to observed red supergiants in the LMC.
\label{fig3}}
\end{figure}

From Fig. \ref{fig2}, the suggestion that sources brighter than $V \sim$ -8.5 (marked with a line) are clusters seems reasonable. It is important to remember, however, that this does not mean that all objects fainter than this magnitude are single sources. It is possible that the population of blue sources contains several early type supergiants, and that the progenitor of SN 2009kr was originally one of these objects, which was transitioning between a blue and a red supergiant when it exploded. A further intriguing possibility is that, despite the apparent colour, the progenitor of this supernova was in fact a luminous blue variable (LBV). Smith et al. \cite{Smi04} have suggested that the bi-stability jump observed at a temperature of $\sim 21000$ K, where the stellar wind properties change from a fast wind, with a low \.{M} to a slow wind with a high \.{M}, may also lead to the formation of a pseudo-photosphere. If this occurs, then an early B star could appear to be a yellow supergiant from its position on the HR diagram. In Smith et al.'s models, the effect is stronger for slightly lower masses (10 $M_\odot$) than that which we find for our progenitor (15 $M_\odot$). Furthermore, the models used by these authors are for higher luminosities, log $L/L_\odot$ = 5.7, than that of our progenitor. We also fail to see the indications of strong circumstellar interaction in the SN spectrum, which we would expect to see for an object which has undergone significant episodes of recent mass loss, such as an LBV. 

Smartt et al. \cite{Sma09b} have suggested that there may be a missing progenitor population of red supergiants (RSG) compared to what one expects from a typical initial mass function and Local Group RSG populations. There have been no detections of RSG SN progenitors above $\log L/L_{\odot} \simeq 5.1$, and one possibility is that stars above this traverse back to the blue and become Wolf-Rayet stars. As the progenitor of SN 2009kr was of this luminosity, is possible that it was in this bluewards moving phase (as was suggested for SN 2008cn by Elias-Rosa et al. \citealp{Eli09}). However this scenario does not sit comfortably with the low metallicity estimate of around 8.1 dex for the SN environment. Some authors have suggested that the classical LBV stage could even stretch down to luminosities as low as 5.1 dex (Smith et al. \citealp{Smi09}). Hence the apparently hotter photosphere than for an RSG could be due to a post-RSG LBV stage during which the core collapses much earlier than most theoretical models predict. 

The progenitor appears to be a single source, although a cluster cannot be ruled out, except by observing the disappearance of the progenitor several years hence (e.g. Gal-Yam \& Leonard \citealp{Gal09}; Maund \& Smartt \citealp{Mau09}). The progenitor of SN 2004et was originally proposed to be a yellow supergiant (Li et al. \citealp{Li05}) of similar luminosity to the progenitor detected here. However, Crockett et al. (2009, in preparation) show that this object was not a single star, but a blend of three sources. The spatial resolution of the NGC 6946 images (the galaxy in which SN 2004et exploded, 0.8$\arcsec$ at 5.9 Mpc) of 23 pc is similar to the resolution of the WFPC2 images. A binary system could also mimic the colour of the progenitor candidate, with a massive, early type star as a companion to a RSG which exploded. At the distance of NGC 1832, even a wide, non-interacting binary could appear as a single source. When the SN fades, future ACS or WF3 images will determine the nature of the source.

\begin{figure}
\begin{center}
\includegraphics[angle=270,scale=0.6]{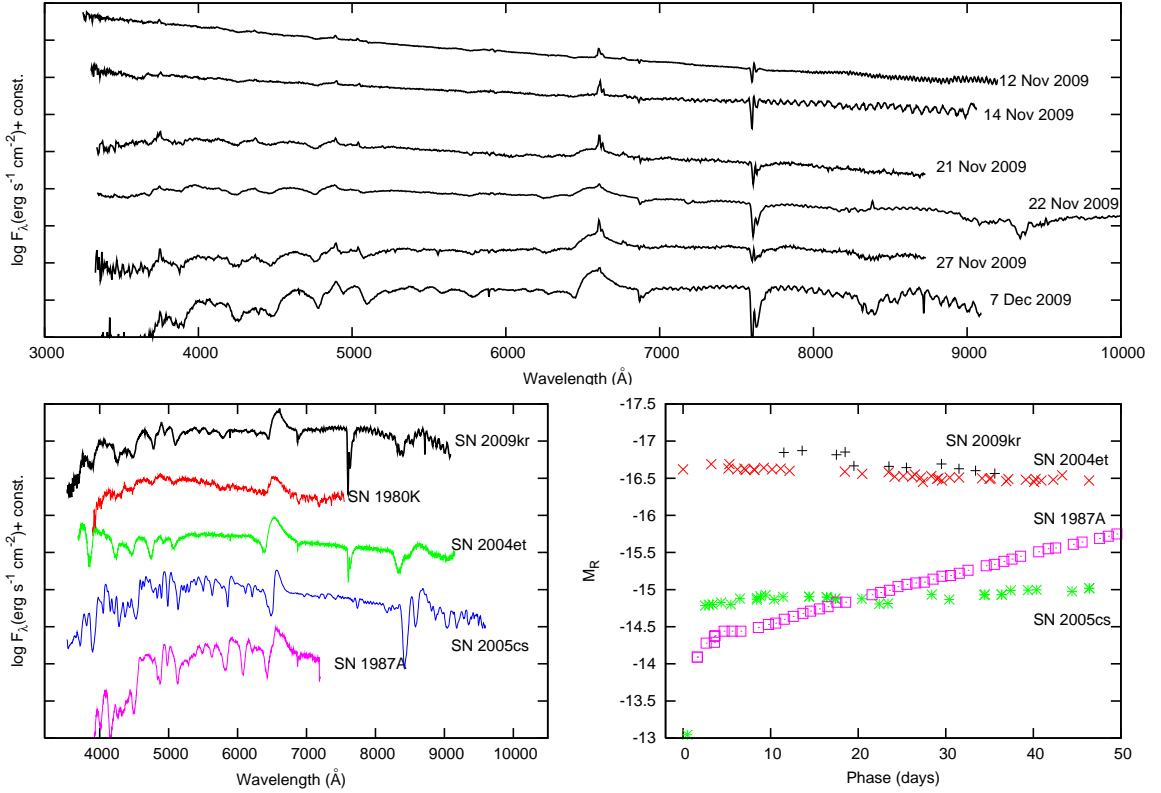}
\caption{Top panel: Time series of spectra of SN 2009kr from Telescopio Nazionale Galileo + Dolores, Nordic Optical Telescope + ALFOSC, Calar Alto 2.2m telescope + CAFOS, New Technology Telescope + EFOSC2, showing the evolution of the SNe, and in particular the development of the $H\alpha$ P-Cygni profile. Narrow components are believed to be from a nearby HII region. Bottom left panel: Spectrum of the SN at $\sim$30d, compared to other SNe at similar epochs, as discussed in text. Bottom right panel: R-band lightcurve of SN 2009kr based on photometry obtained with Liverpool Telescope + RatCam, as compared to other SNe (r$\arcmin$ images from RatCam have been calibrated to Bessell R). SN 2009kr magnitudes are relative to discovery epoch (Nov. 6).
\label{fig4}}
\end{center}
\end{figure}

A sequence of spectra spanning a period of about 1 month is shown in Fig. \ref{fig4} (top panel). Early-time spectra show a very blue continuum and weak lines of H and He I. Later, with the H envelope recombination onset, the continuum becomes redder and metal lines (especially Fe II and Sc II) become prominent. In the middle panel a $\sim$30 d spectrum of SN 2009kr is compared with spectra of SNe 1987A, 2005cs and 2004et at a similar phase. The spectrum of SN 2009kr shows a relatively red continuum and the classical Fe II lines that are ususally visible in Type II SNe during recombination. However, the weakness of Ca II lines in this phase, and the two-component absorption profile of H$\alpha$ are puzzling. The latter can be explained with a very prominent high velocity ($\sim$ 15000 km s$^{-1}$) H component in addition to the canonical moderate-velocity ($\sim$ 7300 km s$^{-1}$) absorption. Alternatively, the bluer shoulder can be interpreted as due to line blending (e.g. with N II). Similar profiles have been already observed in SNe IIP (e.g. in SN 1999em, see Baron et al. \citealp{Bar00}; Dessart \& Hiller \citealp{Des06}), but always with the redder component dominating over the bluer. The dominance of emission over absorption in H$\alpha$ is reminiscent of Type II Linear SNe, and would support the stripped (to some extent) envelope progenitor.

The R-band absolute light curve of SN 2009kr spanning a period of $\sim 35$ days after the core-collapse is shown in Fig. \ref{fig4} (Bottom panel), and is compared with the light curves of three Type II SNe with well-constrained progenitor information: SN 1987A (Menzies et al. \citealp{Men87}), SN 2005cs (Pastorello et al. \citealp{Pas06}, \citealp{Pas09}) and SN 2004et (Maguire et al. 2009). The flat light curve of SN 2009kr suggests that this object is similar to Type IIP SNe. Its light curve is only marginally brighter than that of SN 2004et and it is about 2 mags more luminous than SN 2005cs. The plateau in SNe IIP is attributed to the recombination of the cooling H envelope. SN 1987A (e.g. Arnett \& Fu \citealp{Arn89}) did not show a proper plateau, but a delayed and broad light curve maximum attributed to the small progenitor radius at the time of the explosion. The progenitor of SN 1987A was indeed a compact blue supergiant (Gilmozzi et al. \citealp{Gil87}), whilst more canonical Type IIP SNe are well-known to have red supergiant progenitors (e.g. Smartt \citealp{Sma09a}). In this context, a relatively compact yellow supergiant progenitor like that of SN 2009kr is expected to produce a plateau in the light curve that slowly increases its brightness with time, although without reaching the extreme behaviour of the light curve of SN 1987A. This behavior, however, is not observed in the light curve of SN 2009kr until phase $\sim$ 35 days, although further observations are necessary to establish if it will continue to evolve like a normal SN IIP.

\acknowledgments
This work was conducted as part of a EURYI scheme award (esf.org/euryi). Data were obtained from the Multimission Archive at the Space Telescope Science Institute, operated by the Association of Universities for Research in Astronomy, Inc., under NASA contract NAS5-26555; the NASA/IPAC Extragalactic Database (NED) and the Isaac Newton Group Archive maintained by the CASU at the Institute of Astronomy, Cambridge, and observations made with the Italian Telescopio Nazionale Galileo (TNG) operated on the island of La Palma by the Fundaci—n Galileo Galilei of the INAF (Istituto Nazionale di Astrofisica) at the Spanish Observatorio del Roque de los Muchachos of the Instituto de Astrofisica de Canarias. We thank J.S. Vink for his suggestion of a pseudo-photosphere as a possible explanation for the yellow colour of the progenitor. MF acknowledges funding by the Department of Employment and Learning NI, and SM by Academy of Finland (Project 8120503). KT has received support from Hungarian OTKA Grant K76816 and from the Hungarian E\"otv\"os Fellowship. ST acknowledges support by the Transregional Collaborative Research Centre TRR 33 ``The Dark Universe'' of the German Research Foundation (DFG). 

{\it Facilities:} \facility{HST (WFPC2)}, \facility{VLT:Yepun (NACO)}, \facility{TNG (Dolores)}, \facility{NOT (ALFOSC)}, \facility{CAO:2.2m (CAFOS)}, \facility{NTT (EFOSC)}

\clearpage

\clearpage

\clearpage

\clearpage

\clearpage

%


\end{document}